# IMPLEMENTASI dan ANALISIS PERFORMA BONDING INTERFACE MODE 802.3ad sebagai LINK REDUNDANCY pada ROUTER MIKROTIK

*Implentation and Analysis of Bonding Interface Mode 802.3ad as Link Redundancy on Mikrotik Router*

[1]**Raditya Muhammad**         [2]**Muhammad Iqbal**         [3]**Ratna Mayasari**

[1]Program Studi Rekayasa Perangkat Lunak – Universitas Pendidikan Indonesia
[2,3]Universitas Telkom
Bandung, Indonesia
[1]radityamuhammad@upi.edu

**ABSTRAK**

Kestabilan dan kehandalan jaringan, merupakan syarat mutlak bagi jaringan telekomunikasi. Bonding interface adalah salah satu teknik yang dapat memfasilitasi jaringan untuk melayani layanan-layanan yang membutuhkan kestabilan dan kehandalan jaringan.

Sistem bonding interface ini menggabungkan dua *interface* menjadi sebuah *link virtual* yang ditandai dengan penggunaan satu alamat IP. Apabila suatu frame dikirimkan dari pengirim menuju penerima, namun pada proses pengiriman salah satu *link* tersebut terjadi gangguan, maka *link* yang masih tersambung akan mampu untuk menjaga koneksi agar pengiriman frame tetap berlangsung, sistem kerja ini dinamakan *link redundancy*. Untuk memantau kondisi *link* apakah dalam keadaan tersambung atau putus, digunakan sebuah mekanisme yang dinamakan *link-monitoring*. *Link monitoring* yang digunakan dalam tugas akhir ini adalah Media Independent Interface (MII). Langkah pengujian yang dilakukan yaitu dengan cara mengimplementasikan jaringan Bonding Interface untuk penggunaan *video streaming*, VoIP, dan *file transfer*. Analisis yang dilakukan adalah dengan uji *failover* yang menggunakan parameter waktu pengalihan koneksi dan uji QoS dengan parameter *packet loss*, *delay*, *jitter*, dan *throughput*. Yang kemudian hasilnya akan dibandingkan dengan jaringan yang tidak menggunakan Bonding Interface, yaitu jaringan yang hanya dihubungkan dengan menggunakan satu *link*.

Dari hasil penelitian yang dilakukan adalah bahwa sistem Bonding Interface mampu melakukan mekanisme redundansi saat salah satu *link* putus/*down* ke link yang masih aktif/tersambung dan layanan-layanan yang dijalankan tidak mengalami gangguan. Bonding Interface juga mampu memberikan kestabilan jaringan dibandingkan dengan jaringan yang tidak menggunakan Bonding Interface hal ini ditandai dengan nilai *jitter* yang lebih kecil dibandingkan penggunaan jaringan dengan satu *link*.

Kata Kunci*: Bonding Interface, Link Redundancy, Link monitoring*.

*Abstract*

*Network stability and reliability is an absolute requirement for telecommunications networks. Bonding interface is a technique that can facilitate the network to serve services that require network stability and reliability.*

*This interface bonding system combines two interfaces into a virtual link that is characterized by the use of one IP address. If a frame is sent from the sender to the receiver, but in the process of sending one of the links there is an interruption, then the link that is still connected will be able to maintain the connection so that frame transmission continues, this working system is called link redundancy. To monitor the condition of the link whether it is connected or broken, a mechanism called link-monitoring is used. The monitoring link used in this final project is Media Independent Interface (MII). The test step is to implement a Bonding Interface network for the use of video streaming, VoIP, and file transfer. The analysis carried out is a failover test that uses a connection switching time parameter and a QoS test with packet loss, delay, jitter, and throughput parameters. The results will then be compared with a network that does not use a Bonding Interface, which is a network that is only connected using one link.*

*From the results of the research conducted is that the Bonding Interface system is able to perform a redundancy mechanism when one of the links is broken/down to an active/connected link and the services that are run are not interrupted. Bonding Interface is also able to provide network stability compared to networks that do not use a Bonding Interface, this is characterized by a smaller jitter value than the use of a network with one link.*

***Keywords: Bonding Interface, Link Redundancy, Link monitoring.***

## 1. Pendahuluan
### 1.1 Latar Belakang

Asosiasi Penyelenggara Jasa Internet Indonesia (APJII) mengungkapkan bahwa total pengguna internet di seluruh Indonesia pada tahun 2012 telah mencapai ± 63 juta orang atau 24,23% dari total rakyat Indonesia (Kompas.com:2012). Kebutuhan akan komunikasi merupakan kebutuhan dasar setiap manusia. Oleh karena itu kestabilan dan kehandalan koneksi merupakan faktor utama keberhasilan suatu jaringan.

Salah satu cara untuk meningkatkan kestabilan dan kehandalan koneksi yaitu dengan menggunakan algoritma pemetaan. Algoritma pemetaan dapat mendeteksi dan memilih jalur yang terbaik agar komunikasi tetap berlangsung walaupun ada salah satu perangkat jaringan yang hilang koneksi. namun, menggunakan algoritma pemetaan membutuhkan waktu yang lama untuk pengalihan koneksi dan membutuhkan biaya tambahan untuk penyediaan *router* tambahan.

Oleh karena itu dibuatlah suatu sistem *link redundancy* yang dapat mengatasi masalah putusnya media transmisi. Dengan adanya sistem *link redundancy* ini, koneksi akan lebih terjaga keberlangsungannya karena terdapat mekanisme *failover*.

Sistem *link redundancy* yang dimaksud adalah Bonding Interface. Bonding Interface adalah suatu sistem yang dapat menggabungkan dua atau lebih interface menjadi sebuah *link virtual*, intinya adalah, beberapa interface ini akan memiliki alamat IP (Internet Protocol) yang sama.

Dengan sistem Bonding Interface ini, kestabilan dan kehandalan koneksi terus terjaga, sehingga layanan-layanan yang membutuhkan kestabilan dan kehandalan koneksi jaringan dapat terlayani secara optimal.

### 1.2 Tujuan
Tujuan dari penelitian ini adalah untuk mengetahui dan menganalisis bagaimana sistem kerja Bonding Interface dalam mekanisme *failover,* dan performa jaringan Bonding Interface, dengan dibandingkan dengan jaringan yang tidak menggunakan Bonding Interface.

### 1.3 Perumusan Masalah
Permasalahan yang dibahas adalah bagaimana merancang sebuah jaringan Bonding Interface yang dapat memfasilitasi layanan-layanan yang membutuhkan kestabilan dan kehandalan jaringan dan pengaruh Bonding Interface terhadap kualitas jaringan.

Penelitian yang dilakukan adalah dengan membangun jaringan Bonding Interface menggunakan dua *link* kemudian pada jaringan dilewatkan layanan video, suara, dan data. Pengukuran dilakukan dengan parameter *downtime* dan parameter QoS seperti: *delay, jitter, throughput,* dan *packet loss.*

### 1.4 Batasan Masalah

Dalam penelitian yang dilakukan, permasalahan dibatasi agar pembahasan lebih terfokus, yaitu:
- Jumlah *Interface* yang dibuat untuk sistem bonding adalah dua.
- Analisa yang dilakukan fokus pada bagaimana sistem Bonding Interface digunakan sebagai *link redundancy* dan performanya dalam menentukan kualitas jaringan.
- Tidak membahas spesifikasi perangkat dan layanan secara rinci.

### 1.5 Metode Penelitian
Metode yang digunakan dalam penyusunan tugas akhir ini menggunakan metode studi literatur dan pengukuran empiris.
- Identifikasi Masalah:. Permasalahan yang diidentifikasi adalah Bonding Interface dan pengaruhnya terhadap kualitas jaringan.
- Menentukan tujuan dan manfaat: ditentukan tujuan agar penelitian yang dilakukan memiliki nilai arti yang jelas.
- Studi literatur: pencarian informasi yang dibutuhkan mengenai Bonding Interface melalui berbagai sumber seperti, buku, jurnal, artikel di situs web dll. Adapun daftar media kepustakaan yang dimanfaatkan tercantum dalam daftar pustaka.
- Perancangan sistem: merancang sebuah sistem yang dapat menghasilkan solusi dari permasalahan yang ada.
- Implementasi sistem: dilakukan implementasi sistem yang mengacu kepada desain rancangan sistem yang telah dibuat sebelumnya.
- Pengujian dan Evaluasi: dilakukan pengujian dan pengukuran terhadap sistem yang telah dibuat dan dilakukan evaluasi dengan harapan dapat memperbaiki kesalahan yang telah dibuat

## 2. Landasan Teori
### 2.1 Bonding Interface
Secara bahasa *Bonding Interface* terdiri dari dua suku kata, *Bonding* berarti mengikat dan *Interface* berarti penghubung, maka Bonding Interface adalah ikatan dari suatu penghubung. Bonding Interface adalah sistem yang memungkinkan penggunanya untuk mengikat/menggabungkan beberapa *interface* seperti ethernet menjadi sebuah *link virtual* tunggal [6]. Bonding Interface juga dikenal dengan istilah *Link Aggregation*.

### 2.1.2 Kelebihan Bonding Interface [5]
Bonding Interface memiliki beberapa kelebihan dibandingkan dengan jaringan tanpa menggunakan Bonding Interface, yaitu:
- Untuk mendapatkan kapasitas penyaluran data yang lebih baik : karena pada Bonding Interface menggabungkan dua atau lebih *link* menjadi satu *link* maka kapasitas penyaluran datanya bertambah.
- Kemampuan untuk *link redundancy* : karena ada sistem *link monitoring* maka apabila salah satu *link* terputus maka proses pendistribusian frame akan dialihkan ke *link* lain.

- Kemampuan untuk *load balancing*: trafik dari MAC (Media Access Control) *client* didistribusikan ke beberapa *link* yang di *bonding*.

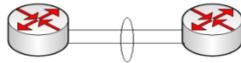

**Gambar 2.1 Bonding Interface**

**2.2  802.3ad [5]**

802.3ad adalah standar yang dikeluarkan oleh Institute of Electrical and Electronics Engineers (IEEE) pada tanggal 30 Maret 2000, standar 802.3ad merupakan bagian dari kelompok standar untuk jaringan area lokal (LAN) dan jaringan area metropolitan (MAN). 802.3ad merupakan standar yang khusus membahas mengenai *link Aggregation full duplex*, *point to point*, dan memiliki *data rate* yang sama (10mb/s, 100mb/s atau 1000mb/s).

Standar IEEE 802.3ad disebut juga sebagai *Link Aggregation Control Protocol* (LACP). Pada layer OSI LACP berkedudukan pada layer Datalink. Protokol ini mengharuskan bahwa pada kedua ujung *link* LACP harus diaktifkan terlebih dahulu. LACP menyediakan fitur kontrol untuk penambahan atau pengurangan *link* tanpa harus ada frame yang hilang saat pengiriman.

Agar LACP dapat menentukan apakah suatu kumpulan *link* dapat diikat atau digabung, ada beberapa kebutuhan yang harus dipenuhi yaitu:
- System ID : suatu pengidentifikasi yang unik yang terdiri dari MAC Prefiks yang dapat digunakan untuk menentukan prioritas sistem.
- Aggregation ID : setiap Aggregator (terdiri dari frame distributor dan frame collector) membutuhkan alamat MAC yang unik, alamat MAC yang digunakan adalah alamat MAC yang sama untuk membuat System ID
- *Port* ID : *Port* ID terdiri dari *port* prioritas yang digabung dengan *port* number.
- LAG ID : *Link* Aggregation Group Id dapat dibentuk dari: System ID, Kunci aggregasi operasional untuk semua *port* yang digabung, dan *Port* ID.

**2.2.1  Mekanisme pendistribusian frame**

Mekanisme pendistribusian frame tergantung terhadap algoritma trafik yang digunakan, algoritma trafik ini harus memastikan bahwa:
- Tidak adanya kesalahan pemesanan frame dalam proses komunikasi
- Tidak ada duplikasi frame yang dikirimkan.

Persyaratan terhadap algoritma trafik tersebut harus dipenuhi agar seluruh frame yang akan dikirim benar-benar akan dikirimkan melalui *link virtual* secara berurutan yang dihasilkan oleh MAC *Client*. Selain itu tidak diperbolehkan adanya penambahan dan pengurangan (modifikasi) pada frame yang akan dikirimkan.

Frame-frame yang dikirim oleh MAC *client* diambil oleh distributor, selanjutnya dikirimkan melalui *port* yang sesuai, selanjutnya dipenerima akan diterima melalui *port* yang sesuai lalu, dilewatkan frame yang akan dikirimkan ke kolektor untuk disampaikan ke MAC *client* tujuan.

**2.4  Mikrotik**

Mikrotik berasal dari kata "*mikrotikls*" bahasa Latvia yang berarti "jaringan kecil". Mikrotik berkantor pusat di Latvia, Produk-produk *hardware* yang merupakan produksi unggulan Mikrotik yaitu: *router*, *switch*, antena, dan perangkat pendukung lainnya, sedangkan untuk produk *Software* unggulan Mikrotik adalah MikroTik *Router*OS.

*Router*Board (RB) adalah produk *router* dari Mikrotik yang telah tertanamkan sistem operasi *Router*OS. *Router*Board seperti sebuah PC mini yang terintegrasi karena dalam satu *board* tertanam prosesor, RAM, ROM, dan memori flash. *Router*Board berfungsi sebagai *router* jaringan, *bandwidth management*, proxy *server*, DHCP, DNS *server* dan bisa juga berfungsi sebagai *hotspot server*.

**2.5  VoIP**

Pada dasarnya teknologi VoIP ini bekerja dengan cara merubah format suara manusia yang analog, namun akan dikirimkan menjadi format data digital tertentu dengan menggunakan proses coder dan decoder, selanjutnya dikirimkan melalui jaringan berbasis IP. Untuk proses signalling, terdapat dua standar protokol yang digunakan, yaitu: H323 dan *Session Initiation Protocol* (SIP). H323 adalah protokol yang dikeluarkan ITU-T sedangan SIP adalah protokol yang dikeluarkan oleh IETF. Proses signaling menggunakan H323 menumpang pada protokol trans*port* TCP, sedangkan SIP pada TCP dan UDP. Protokol yang bertanggung jawab untuk media trans*port* adalah *Real-Time Transport Protocol* (RTP).

**2.6  *Streaming***

*Streaming* adalah teknologi jaringan *client-server* untuk memainkan *file* audio/video yang terletak pada *server* secara langsung diputar pada *client* sesaat setelah ada permintaan dari *client*.

Saat *file* di-*stream* maka akan terbentuk *buffer* di *client*. Data audio/video akan diunduh ke dalam *buffer*. Setelah *buffer* terisi dalam beberapa detik kemudian secara otomatis *file* video/audio akan di jalankan oleh sistem. Sistem akan membaca informasi dari *buffer* sambil tetap melakukan proses unduh *file* sehingga proses *streaming* tetap berlangsung ke mesin *client*.

**2.7  *File Transfer***

*File transfer* adalah proses komunikasi data dari suatu perangkat ke perangkat lainnya. Protokol yang mengatur adalah *File Transfer Protocol* (FTP). FTP adalah protokol standar untuk proses komunikasi data antar suatu perangkat dengan perangkat lainnya dalam satu jaringan ataupun berbeda jaringan.

**3.  Perancangan Dan Implementasi Sistem**

**3.1  Perancangan Sistem**

Pada proses implementasi digunakan empat buah *router* untuk menjalankan sistem ini. Pada masing-masing *router* dihubungkan melalui dua buah *port* yang menjadi sebuah *link virtual* Bonding Interface. Keempat *router* tersebut dihubungkan dengan pola pe*routing*an OSPF. Pengujian yang dilakukan, yaitu pada jaringan dilewatkan tiga buah layanan yang berbeda format, yaitu VoIP (suara), Video *Streaming* (Video), dan *File Transfer* (Data). Pada penelitian terdapat dua skenario yang akan dilakukan, yaitu pengujian *Failover* dan pengujian *Quality of Service* (QoS).

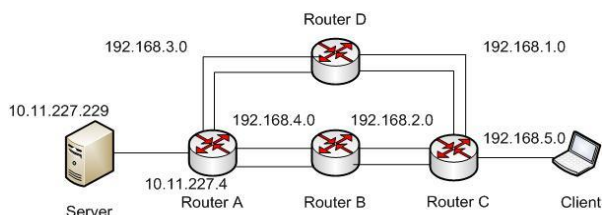

**Gambar 3.1 Topologi Jaringan**

### 3.2 Pengujian Sistem
#### 3.2.1 Pengujian Failover

Pengujian dilakukan untuk menganalisis waktu yang dibutuhkan untuk mengalihkan trafik dari *link* yang putus ke *link* yang masih tersambung. Pengujian yang dilakukan, pada jaringan dilewatkan layanan video, suara dan data secara bersamaan, hal ini dimaksudkan untuk memberikan skema trafik yang padat pada jaringan, kemudian dilewatkan paket uji sebesar 32 byte menggunakan Ping. Saat trafik sedang berjalan dilakukan dua langkah pengujian, yaitu:
1. Mencabut satu *link* pada *router*.
2. Mencabut dua *link* sekaligus pada *router*.

Pada pengujian ini digunakan parameter *downtime*,

#### 3.2.2 Pengujian QoS

Pengujian dilakukan untuk menganalisis performa jaringan Bonding Interface dan *Non-Bonding* Interface saat dilewatkan layanan data, suara dan video, dan kemudian dibandingkan masing-masing perfomansinya. Topologi jaringan *Non-Bonding* Interface sama, namun setiap *router* hanya tersambung oleh satu *interface* dan protokol pemetaan yang digunakan adalah OSPF.

Pengujian yang dilakukan adalah pada jaringan Bonding Interface dan jaringan yang tidak menggunakan Bonding Interface, akan dilakukan dua kondisi yaitu:
1. Dilewatkan layanan data, suara, dan video secara bergantian
2. Dilewatkan layanan data, suara, dan video secara bersamaan

parameter yang digunakan pada pengujian yaitu: *throughput*, *delay*, *jitter* dan *packet loss*.

### 4. Analisis Sistem
### 4.2 Pengujian *Failover*

Fungsi utama dari Bonding Interface adalah menyediakan mekanisme *failover*. *Failover* adalah suatu teknik/metode pengalihan trafik dari *link* yang putus ke *link* yang masih tersambung, dengan tujuan agar komunikasi dapat terus berlangsung. Pada pengujian *failover* digunakan parameter *downtime*, *downtime* adalah waktu yang menyatakan saat suatu perangkat di dalam jaringan tidak terkoneksi dengan perangkat lainnya, tetapi komunikasi dapat tetap berlangsung karena terdapat perangkat lain yang menjaga komunikasi tetap berlangsung.

Pengujian dimulai dengan menjalankan *streaming* video, VoIP dan unduh *file* dari *server* secara bersamaan, kemudian ping menggunakan Mikrotik, Saat satu *link* dicabut. Waktu *downtime* diukur dengan cara mengalikan jumlah *timeout* dengan 20 (nilai patokan *timeout*). Sedangkan untuk pengujian dua buah *link* yang dicabut sekaligus diukur dari ping yang berada pada cmd di komputer *client* dan pengukuran menggunakan *software network protocol analyzer* Wireshark.

**Tabel 4.1 Hasil Pengukuran *Downtime* Ping Ke Arah *Server* (dalam ms)**

| Jaringan | IP Server | IP Gateway Server |
|---|---|---|
| 192.168.1.0 | 358,667 | 368,000 |
| 192.168.2.0 | 326,667 | 372,667 |
| 192.168.3.0 | 395,333 | 392,000 |
| 192.168.4.0 | 360,000 | 414,667 |

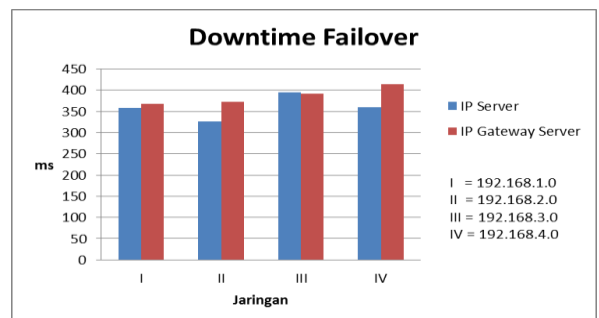

**Gambar 4.1 *Downtime* Ping Ke Arah *Server***

*Downtime* yang terjadi saat dilakukan ping ke arah *server* dari *router* yang diamati menunjukkan bahwa waktu *downtime* saat ping ke gateway *server* lebih lama dibandingkan saat ping ke komputer *server*, hal ini disebabkan karena waktu Time-To-Live (TTL) yang ditentukan oleh Mikrotik untuk pengiriman paket dengan tujuan *router* lebih kecil (63ms/64ms) dibandingkan dengan TTL untuk dengan tujuan komputer (126ms/127ms).

**Tabel 4.2 Hasil Pengukuran *Downtime* Ping Ke Arah *Client* (dalam ms)**

| Jaringan | IP Client | IP Gateway Client |
|---|---|---|
| 192.168.1.0 | 321,333 | 525,333 |
| 192.168.2.0 | 355,333 | 519,333 |
| 192.168.3.0 | 290,000 | 439,333 |
| 192.168.4.0 | 325,333 | 419,333 |

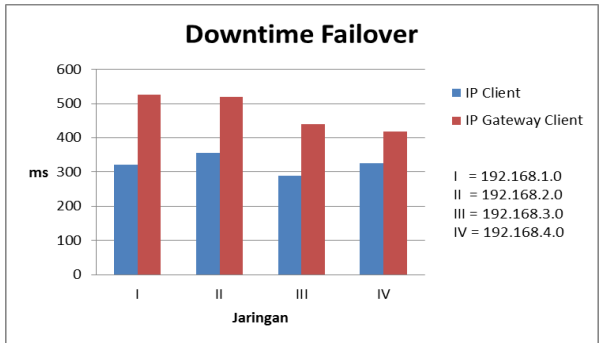

**Gambar 4.2** *Downtime* Ping Ke Arah *Client*

Sama halnya dengan *Downtime* yang terjadi saat dilakukan ping ke arah *server*, dari *router* yang diamati menunjukkan bahwa waktu *downtime* saat ping ke gateway *client* lebih lama dibandingkan saat ping ke komputer *server* hal ini disebabkan karena waktu TTL yang ditentukan oleh Mikrotik untuk pengiriman paket dengan tujuan *router* lebih kecil (63ms/64ms) dibandingkan dengan TTL untuk dengan tujuan komputer (126ms/127ms).

**Tabel 4.3 Hasil Pengukuran *Failover* dengan Jumlah *Link* Berbeda**

| Jaringan | Satu Link | Dua Link |
|---|---|---|
| 192.168.1.0 | 0,359 detik | 15,712 detik |
| 192.168.2.0 | 0,327 detik | 37,493 detik |

Dari Tabel 4.3 terlihat bahwa waktu saat sebuah jaringan berada dalam keadaan down/putus hingga jaringan tersebut terkoneksi kembali terjadi perbedaan waktu sebesar 15,353 detik pada jaringan 192.168.1.0 dan 37,166 detik pada jaringan 192.168.2.0, hal ini dapat terjadi karena pada pemutusan dua *link* sekaligus, mekanisme yang bekerja adalah mekanime pemetaan OSPF, waktu yang dibutuhkan untuk pemutusan dua *link* menunjukkan waktu proses saat *router* kembali meng*update* tabel pemetaan jaringannya, baru kemudian kembali dipetakan jalur yang dapat digunakan untuk proses pengiriman paket.

### 4.3 Pengujian QoS

Pengujian dilakukan dengan menjalankan layanan video dengan menggunakan aplikasi VLC 2.1.0 dengan menggunakan transcoding MPEG-2 + MPGA(TS) dengan codec video menggunakan MPEG-2 yang memiliki bitrate 800kb/s dan untuk audio menggunakan codec MPEG Audio dengan bitrate 128kb/s dan *samplerate* 44100Hz, besar video adalah 465MB dan format video "mkv".

Pada VoIP hubungan komunikasi menggunakan softphone 3CX versi 6.0.20943 dengan protokol SIP, dilakukan pemanggilan dari komputer *server* yang bertindak sebagai client 2 ke komputer client, codec yang digunakan berjenis PCM yang dispesifikasikan oleh ITU-T menjadi G.711/ulaw 8bit, memiliki samplerate 8000 sample/detik yang , masing-masing sample tersebut pada codec dikodekan dalam waktu 20ms. Satu frame terdiri dari 160byte. Pada pengukuran one-way *delay*, *delay* yang terukur pada aplikasi Wireshark adalah *propagation delay*, sehingga untuk *one-way delay* perlu ditambahkan dengan *processing delay* 10ms, *algorithmic delay* 3,75ms, *packetization delay* 20ms, *serialization delay* 16,3µsec, de-commmpression *delay* 1ms, de-*packetization delay* 20ms, dan *de-jitter buffer delay* 45ms [2]. Untuk data dilakukan unduh data dari *server* dengan besar data sebesar 699MB.

#### 4.3.1 Hasil Pengujian Layanan Bersamaan

Hasil yang diperoleh saat dilakukan pengujian dengan menjalankan layanan video dengan ukuran video 465MB dengan format video .mkv, VoIP dan data sebesar 699MB secara bersamaan.. Pengujian dilakukan sebanyak 5 kali dengan rentang waktu selama 30 detik.

**Tabel 4.4 Analisis Hasil Pengujian Layanan Secara Bersamaan Pada layanan Video**

| Kriteria | Delay (ms) | Jitter (ms) | Throughput (Mbps) | Packet loss (%) |
|---|---|---|---|---|
| Bonding Interface | 3,142 | 0,758 | 3,125 | 1,42 |
| Non-Bonding Interface | 2,7 | 2,041 | 5,606 | 0,118 |

pada layanan video, rata-rata *delay* yang dihasilkan oleh Bonding Interface lebih besar dibandingkan dengan Non-Bonding Interface, hal ini dikarenakan pada Bonding Interface terdapat mekanisme yang mengatur trafik akan melewati interface yang mana sehingga menambah waktu pengiriman. Namun pada *jitter* Bonding Interface memiliki nilai yang lebih kecil, hal ini menandakan bahwa frame yang diterima di penerima mengalami kestabilan. *Throughput* pada Bonding Interface memiliki nilai yang lebih kecil karena penentuan link yang akan dilewati membuat frame-frame banyak yang tidak mampu dilayani, sehingga menyebabkan nilai *packet loss* pada Bonding Interface pun lebih besar, hal ini dikarenakan kapasitas *router* yang tidak mampu menangani beban trafik, sehingga fitur Bonding Interface tidak berjalan optimal.

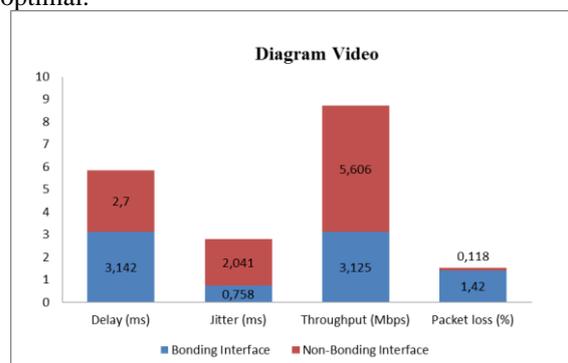

**Gambar 4.3 Diagram Video Penggunaan layanan Bersamaan**

**Tabel 4.5 Analisis Hasil Pengujian Layanan Secara Bersamaan Pada layanan Suara**

| Kriteria | Delay (ms) | Jitter (ms) | Throughput (Mbps) | Packet loss (%) |
|---|---|---|---|---|
| Bonding Interface | 125,796 | 28,319 | 0,131 | 1,422 |
| Non-Bonding Interface | 125,853 | 28,357 | 0,171 | 0 |

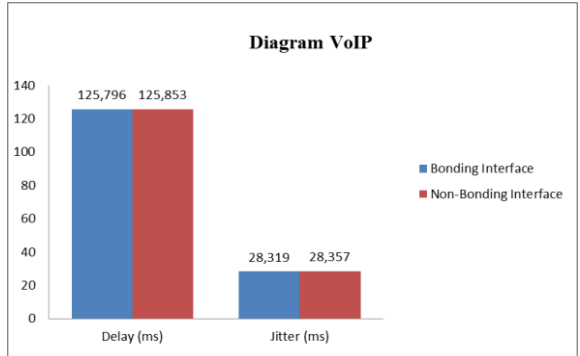

**Gambar 4.4 Diagram Suara Penggunaan layanan Bersamaan (1)**

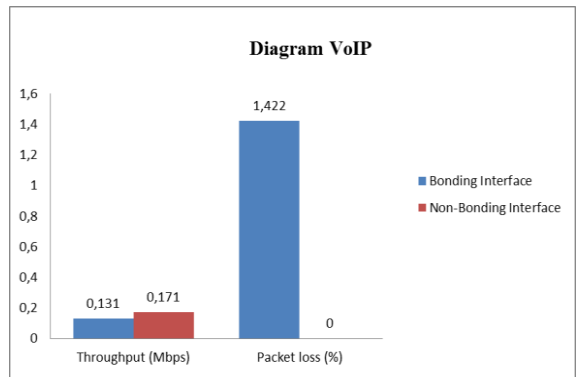

**Gambar 4.5 Diagram Suara Penggunaan layanan Bersamaan (2)**

Pada layanan suara secara umum kualitas jaringan tidak terjadi perbedaan yang jauh, *delay* yang dihasilkan pun rata-rata perbedaan kurang dari 0,5 ms dari jaringan Non-Bonding Interface, hal ini terjadi karena komunikasi dua arah dari *client* 1 dan 2, sehingga kedua *link* aktif bekerja bersama. Pada *Throughput* dan *packet loss* mekanisme penetuan link menyebabkan nilai *throughput* pada Bonding Interface lebih kecil dan nilai *packet loss* lebih besar.

**Tabel 4.6 Analisis Hasil Pengujian Layanan Secara Bersamaan Pada layanan Data**

| Kriteria | *Throughput* (Mbps) |
|---|---|
| Bonding Interface | 85,134 |
| Non-Bonding Interface | 91,696 |

Pada layanan data menunjukkan bahwa penggunaan Bonding Interface tidak mampu memberikan nilai *throughput* yang lebih besar, hal ini dikarenakan kapasitas router yang tidak mampu menangani beban trafik sehingga fitur Bonding Interface tidak berjalan optimal.

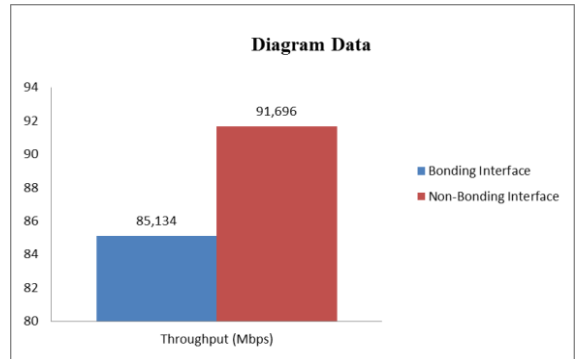

**Gambar 4.6 Diagram Data Penggunaan layanan Bersamaan (2)**

### 4.3.2 Hasil Pengujian Layanan Secara Bergantian

Hasil yang diperoleh saat dilakukan pengujian dengan menjalankan layanan video, suara dan data secara bergantian.

**Tabel 4.7 Analisis Hasil Pengujian Layanan Secara Bergantian Pada layanan Video**

| Kriteria | *Delay* (ms) | *Jitter* (ms) | *Throughput* (Mbps) | *Packet loss* (%) |
|---|---|---|---|---|
| Bonding Interface | 3,801 | 1,178 | 4,205 | 0 |
| Non-Bonding Interface | 3,263 | 1,086 | 4,1 | 0 |

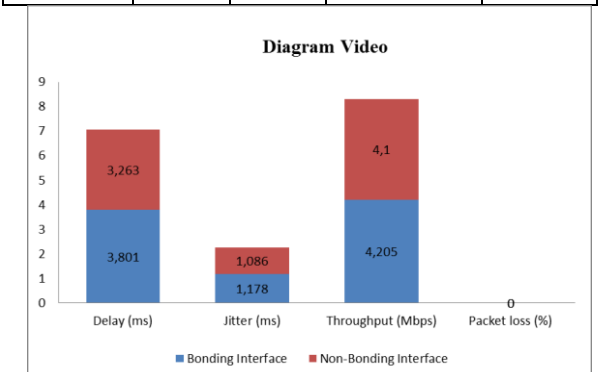

**Gambar 4.7 Diagram Video Penggunaan layanan Bergantian**

Pada layanan video secara umum nilai yang dihasilkan dari tiap-tiap parameter cukup seimbang, karena trafik yang dilewati hanya satu layanan sehingga nilai *throughput* pada Bonding Interface mampu menghasilkan nilai yang lebih besar. Tidak terdapat *packet loss* dikarenakan jaringan mampu untuk melayani setiap paket yang melewati jaringan

**Tabel 4.8 Analisis Hasil Pengujian Layanan Secara Bergantian Pada layanan Suara**

| Kriteria | *Delay* (ms) | *Jitter* (ms) | *Throughput* (Mbps) | *Packet loss* (%) |
|---|---|---|---|---|
| Bonding Interface | 103,567 | 28,319 | 0,262 | 0 |
| Non-Bonding Interface | 103,029 | 28,385 | 0,264 | 0 |

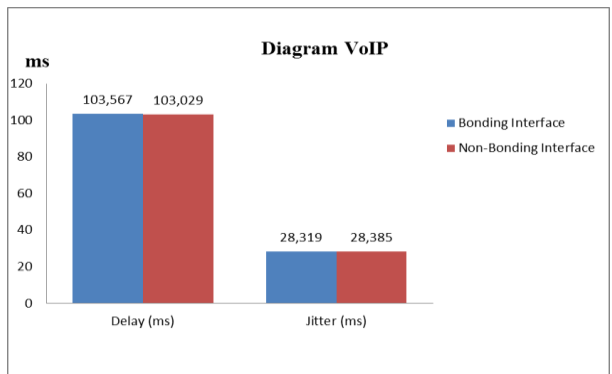

**Gambar 4.8 Diagram Suara Penggunaan layanan Bergantian (1)**

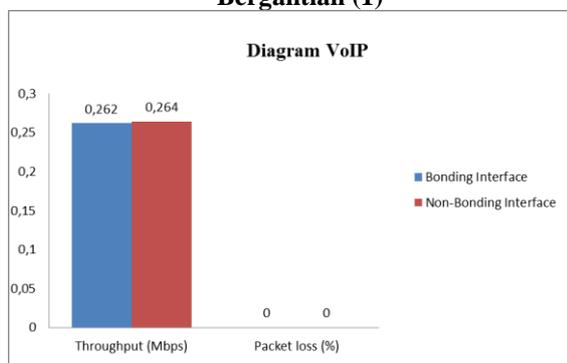

**Gambar 4.9 Diagram Suara Penggunaan layanan Bergantian (2)**

Pada layanan suara secara umum nilai yang dihasilkan dari tiap-tiap parameter cukup seimbang, Tidak terdapat *packet loss* dikarenakan jaringan mampu untuk melayani dengan baik setiap paket yang melewati jaringan.

**Tabel 4.7 Analisis Hasil Pengujian Layanan Secara Bergantian Pada layanan Data**

| Kriteria | *Throughput* (Mbps) |
|---|---|
| Bonding Interface | 90,696 |
| Non-Bonding Interface | 97,591 |

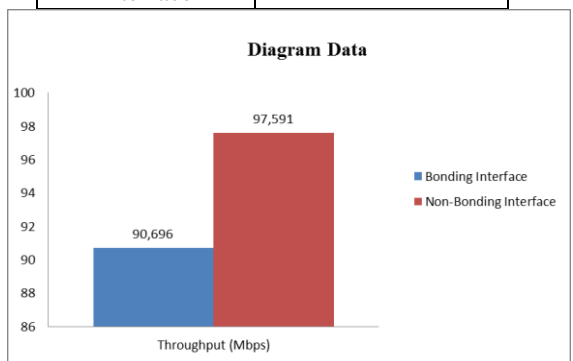

**Gambar 4.10 Diagram DataPenggunaan layanan Bergantian (2)**

Pada layanan data menunjukkan bahwa penggunaan Bonding Interface tidak mampu memberikan nilai *throughput* yang lebih besar hal ini dikarenakan kapasitas router yang tidak mampu menangani beban trafik sehingga fitur Bonding Interface tidak berjalan optimal.

### 5. Kesimpulan dan Saran
**5.1 Kesimpulan**
Berdasarkan tujuan penelitian, proses implementasi, pengujian dan analisis dapat ditarik kesimpulan bahwa:
1. Bonding Interface dapat bekerja sebagai *link redundancy* saat *link* utama dalam keadaan terputus maka *link* lainnya akan menangani (*failover*), sehingga layanan video, suara, dan data tetap berjalan walaupun salah satu *link down*. Penentuan *link* utama oleh protokol LACP adalah berdasarkan tujuan paket akan dikirim. Berdasarkan pengujian yang dilakukan Bonding Interface melakukan *failover* sebesar 386,458ms atau 0,386 detik.
2. Performa Bonding Interface pada keseluruhan mampu memberikan QoS yang sesuai standar. Pada parameter *delay* Bonding Interface terjadi penambahan waktu karena perlu ditentukan *link* yang akan dilalui frame.
3. Berdasarkan pengujian QoS, penggunaan jaringan Bonding Interface tidak meningkatkan performa pada parameter *delay, packet loss dan throughput* dibandingkan dengan jaringan Non-Bonding Interface. Namun Bonding Interface mampu memberikan kestabilan jaringan, hal ini ditandai dengan nilai parameter *jitter* yang lebih kecil dibandingkan dengan jaringan Non-Bonding Interface.

**5.2. Saran**
Saran yang dapat diajukan untuk penelitian selanjutnya adalah:
1. Untuk mengetahui performa jaringan Bonding Interface apakah dapat meningkatkan QoS, perlu dilakukan Bonding Interface dengan jumlah *link* lebih dari dua.
2. Untuk penelitian selanjutnya sebaiknya dilakukan pada media perantara optik agar dapat diterapkan untuk kebutuhan industry jaringan uag lebih cepat.
3. Untuk pengembangan selanjutnya sebaiknya dilakukan pada medium wireless yang dapat menunjang mobilitas *user*.

**DAFTAR PUSTAKA**
[1]. Agung, Rizki.Apa itu Mikrotik, Pengertian, dan Penjelasan. Diakses pada 10 Mei 2013 http://mikrotikindo.blogspot.com/2013/02/apa-itu-mikrotik-pengertian-mikrotik.html
[2]. Cisco."Understanding *Delay* in Packet Voice Networks". Diakses pada 11 Januari 2014 http://www.cisco.com/en/US/tech/tk652/tk698/technologies_white_paper09186a00800a8993.shtml

[3]. Freeman, R. L. 2005."Metropolitan Area Network" in *Fundamentals of Telecommunications.* New Jersey: John Wiley & Sons. pp 341.
[4]. Hard, Dian. 2009. Definisi Bandwidth dan Data Transfer. diakses pada 24 Maret 2013. http://harddian.com/2009/03/18/definisi-bandwidth-dan-data-transfer/#more-73
[5]. IEEE Standard Association. 2000." IEEE Standard for Information Technology - Local and Metropolitan Area Networks - Part 3: Carrier Sense Multiple Access with Collision Detection (CSMA/CD) Access Method and Physical Layer Specifications-Aggregation of Multiple Link Segments. IEEE Computer Society.
[6]. ITT. 2010. Teknologi Streaming. diakses pada 24 Maret 2013 http://digilib.ittelkom.ac.id/index.php?view=article&catid=6%3Ainternet&id=691%3Asteam&tmpl=component&print=1&page=&option=com_content&Itemid=14
[7]. Mikrotik. 2004. *Interfaces Bonding.* diakses pada 2 Maret 2013 http://wiki.mikrotik.com/wiki/Manual:Interface/Bonding.
[8]. Sopandi, D. *Instalasi dan Konfigurasi Jaringan Komputer.* Bandung: Informatika. 2008
[9]. Munadi, Rendy. 2009. "Teknik Switching".Bandung: Informatika. 2009
[10]. Wendi, Raid Indra. Arti dan Fungsi Mikrotik, *Router*Os dan *Router* Board. diakses pada 10 Mei 2013. http://elsyadai.blogspot.com/2012/10/arti-dan-fungsi-mikrotik-*router*os-dan.html
[11]. Tanenbaum, A. S. *Computer Network* (fourth ed.). New Jersey: Prentice Hall PTR. 2003
[12]. ---.2009. Bonding. Linux Foundation. diakses pada 6 Juni 2013 http://www.linuxfoundation.org/collaborate/workgroups/networking/bonding#Configuring_Bonding_Manually